\documentclass[pdflatex,sn-apa]{sn-jnl}

\newtheorem{theorem}{Theorem}

\newtheorem{myDef}{Definition}[section]
\newtheorem{example}{Example}[section]
\usepackage[justification=centering]{caption}
\newtheorem{myproof}{Proof}[section]

\usepackage{lineno,hyperref}
\usepackage{amsmath}
\usepackage{indentfirst}
\usepackage{booktabs}
\usepackage{subfigure}
\usepackage{epsfig}
\usepackage{amssymb}
\usepackage{graphicx}
\usepackage{float}
\usepackage{setspace}
\usepackage{hyperref} 
\usepackage{tabularx,booktabs}
\usepackage{url}
\usepackage{graphicx}
\usepackage{mathrsfs}
\usepackage{enumerate}%

\jyear{2021}%

\begin{document}

\title[Maximum Entropy of Random Permutation Set]{Maximum Entropy of Random Permutation Set}


\author[1]{\fnm{Jixiang} \sur{Deng}}

\author*[1,2,3,4]{\fnm{Yong} \sur{Deng}}\email{dengentropy@uestc.edu.cn}

\affil[1] {Institute of Fundamental and Frontier Science, University of Electronic Science and Technology of China, Chengdu, 610054, China}
\affil[2] {School of Education, Shannxi Normal University, Xi'an, 710062, China}
\affil[3] {School of Knowledge Science, Japan Advanced Institute of Science and Technology, Nomi, Ishikawa 923-1211, Japan}
\affil[4] {Department of Management, Technology, and Economics, ETH Zurich, Zurich, 8093, Switzerland}



\abstract{Recently, a new type of set, named as random permutation set (RPS), is proposed by considering all the permutations of elements in a certain set. For measuring the uncertainty of RPS, the entropy of RPS is presented.  However, the maximum entropy principle of RPS entropy has not been discussed. To address this issue, in this paper, the maximum entropy of RPS is presented. The analytical solution for maximum entropy of RPS and its corresponding PMF condition are respectively proofed and discussed. Numerical examples are used to illustrate the maximum entropy RPS. The results show that the maximum entropy RPS is compatible with the maximum Deng entropy and the maximum Shannon entropy. When the order of the element in the permutation event is ignored, the maximum entropy of RPS will degenerate into the maximum Deng entropy.  When each permutation event is limited to containing just one element, the maximum entropy of RPS will degenerate into the maximum Shannon entropy.}




\maketitle

\section{Introduction}
In order to model and process the information with uncertainty, many theories have been developed, such as probability theory~\citep{lee1980probability}, fuzzy set theory~\citep{zadeh1965fuzzy}, Dempster-Shafer evidence theory (evidence theory)~\citep{dempster1967,Shafer1976}, rough set theory~\citep{pawlak1982rough}, and Z-numbers~\citep{zadeh2011note}, which have been further used in various fields,  such as  information fusion~\citep{lai2020multi,pan2020multi},   risk assessment \citep{8926527,wang2021resilience,gao2021NET},  game theory \citep{cheong2019paradoxical,babajanyan2020energy},     fault diagnosis \citep{Chen2021mleSNPS,huang2021spikingneuralPsystems,wang2020variabletopologies,wang2020electricalsubstations},  complex networks \citep{wen2021fractal}, target recognition \citep{wen2021rotation,xiao2021caftr}, classification \citep{liu2020combination2,liu2020combination1} and decision making~\citep{xie2021informationquality}.

Set theory is a fundamental theory, which provides a basis for most existing theories \citep{jech2013set}.  For example, the sample space of  probability theory is the set that  contains  all the possible outcomes of a certain experiment~\citep{lee1980probability}. In Dempster-Shafer evidence theory, the power set  considers all the possible subsets of a frame of discernment~\citep{dempster1967,Shafer1976}.

Recently, \cite{song2021possible} pointed out that a power set can be viewed as all the possible combination of the elements in a frame of discernment. A straightforward question is: ``what   a set would be if it not consider  all the combinations of elements but all the permutations of elements in the frame of discernment?'' 
To solve this problem, \cite{deng2021RPS} proposed a new type of set, called Random permutation set (RPS), which consists of permutation event space (PES) and permutation mass function (PMF). The PES of a certain set considers all the permutations of that set. PMF describes the chance that a certain element in PES  would happen. 

As is an efficient tool for deal with uncertainty, lots of information entropy have been proposed, such as Shannon entropy \citep{Shannon1948A}, Tsallis entropy \citep{tsallis1988possible,Tsallis2021}, and Deng entropy \citep{deng2020uncertainty,Xue2021Interval}. Also, entropy has  a wide variety of applications, including  effect algebras \citep{di2005entropy},   data fusion \citep{xiao2019multi},  decision making~\citep{xiao2019efmcdm}, and uncertainty measurements \citep{Gao2021Information,zhou2021eXtropy,Xue2021Interval}. For measuring the uncertainty of RPS,  \cite{deng2021RPSentropy} presented the entropy of RPS, which  is compatible with Deng entropy and Shannon entropy.

The maximum entropy principle states that the  distribution with maximal entropy best represents the current state of a system based on given information, which  is widely used in many fields, such as  negation transformation \citep{yager2014maximum,wu2021negation}, imprecise side-conditions \citep{buckley2005maximum}, fuzzy cognitive maps \citep{feng2019learning}, decision making \citep{yager2009weighted}, and uncertainty theory \citep{ma2021remark}. As for the maximum entropy principle of RPS entropy, two aspects should be handled. The first one would be, ``what is the maximum form for the entropy of RPS?'' The second one would be, ``what is the PMF condition for the maximum entropy of RPS?''

To address these issues, in this paper, the maximum entropy of RPS is presented. The analytical solution for maximum entropy of RPS and its corresponding PMF condition are respectively  proofed and discussed.  In addition, numerical examples are used to illustrate the maximum entropy RPS. The results show that  the maximum entropy RPS is compatible with the maximum Deng entropy and the maximum Shannon entropy. When the order of the element in permutation event is ignored, the maximum entropy of RPS will degenerate into the maximum Deng entropy.  When each permutation event is limited to containing just one element, the maximum entropy of RPS will degenerate into the maximum Shannon entropy.

The rest of this article is as follows. Section 2 introduces the preliminaries. Section 3 presents the maximum entropy of RPS. Section 4 uses some numerical examples to illustrated the   maximum entropy of RPS. Section 5 makes a brief conclusion.

\section{Preliminaries}
In this section, we briefly review some preliminaries of this paper.

\subsection{Dempster-Shafer evidence theory}

Dempster-Shafer evidence theory~\citep{dempster1967,Shafer1976} is an efficient tool for dealing with uncertainty~\citep{deng2020uncertainty}.  

\begin{myDef}[Frame of discernment] Let $\Theta$, called the frame of discernment (FOD), denotes a fixed set of $N$ mutually exclusive and exhaustive elements, indicated by $\Theta = \{\theta_1, \theta_2, \ldots, \theta_N\}$. 	The power set of $\Theta$, denoted as $2^\Theta$, contains all possible subsets of $\Theta$ and has $2^N$ elements. $2^\Theta$ is indicated by $2^\Theta  =  \{\varnothing, \{\theta_1\}, \cdots, \{\theta_N\}, \{\theta_1,\theta_2\},  \cdots, \{\theta_1,\theta_N\}, \cdots, \Theta \}$.
\end{myDef}

\begin{myDef} [Mass function] Given a FOD of $\Theta$, a mass function  is a mapping function defined by 
	$m: 2^\Theta \rightarrow [0, 1]$, which is constrained by $m(\emptyset) = 0$ and $\sum\limits_{A \in 2^{\Theta}} m(A) = 1$. 
\end{myDef}

\subsection{Deng entropy and maximum Deng entropy}

In order to measure the uncertainty of mass function in evidence theory, a novel entropy, called Deng entropy, also named as belief entropy, is proposed  \citep{deng2020uncertainty}, which is further generalized into information volume~\citep{deng2020information,deng2021fuzzymembershipfunction,Gao2021Information}.

\begin{myDef}[Deng entropy]
	Given a certain mass function distribution defined on FOD $\Theta$, Deng entropy is defined as:
	\begin{equation}\label{S6}
	H_{DE}(m)=-\sum_{A \in 2^{\Theta}}m(A)\log (\frac{m(A)}{2^{\left\lvert A\right\lvert }-1})
	\end{equation}
	where $\left\lvert  A \right\lvert $ is the cardinal of a certain focal element $A$.
\end{myDef}

The maximum entropy is an important conception in statistic.    The analytical solution of the maximum Deng entropy  and  its corresponding condition are given  as follows \citep{deng2020uncertainty}.

\begin{theorem}[The mass function condition of maximum Deng entropy]
	Let the frame of discernment be $\Theta$.	
	The maximum Deng entropy appears if and only if the mass function distribution satisfies
	\begin{equation}
	m(A)=
	\frac{(2^{\left\lvert A\right\lvert }-1)}{\sum_{A \in 2^{\Theta}}(2^{\left\lvert A\right\lvert }-1)	}, A \in 2^{\Theta}
	\end{equation}
\end{theorem}

\begin{theorem}[The analytic solution of maximum Deng entropy]
	The analytical solution of maximum Deng entropy is that
	\begin{equation}
	H_{\max \text{-}DE}=
	\log \sum_{A \in 2^{\Theta}}(2^{\left\lvert A\right\lvert }-1)
	\end{equation}
\end{theorem}

\subsection{Random permutation set}

In 2021, \cite{deng2021RPS} proposed Random permutation set (RPS), which is a novel set consisting of permutation event space (PES) and permutation mass function (PMF). Some basic definitions of RPS are given as follows.

\begin{myDef}[Permutation event space] Given a fixed set of $N$ mutually exclusive and exhaustive elements $\Theta= \{\theta_1, \theta_2, \ldots, \theta_N\}$, its \textbf{permutation event space (PES)}   is  a set of all  possible permutations of $\Theta$ defined as follows:
	\begin{align}
	PES\left( \Theta \right) =&\{\, A_{ij}\,\lvert \, i= 0, ... , N;\,  j=1, ... ,P(N,i) \, \}\\
	=&\{\,\, \emptyset ,\left( \theta _1 \right) ,\left( \theta _2 \right) , ... ,\left( \theta _N \right) ,\left( \theta _1,\theta _2 \right) ,\left( \theta _2,\theta _1 \right) , ... ,\left( \theta _{N-1},\theta _N \right), \notag\\
	&\,\,\,\left( \theta _{N},\theta _{N-1} \right) , ... ,\left( \theta _1,\theta _2, ... ,\theta _N \right) , ... ,\left( \theta _N,\theta _{N-1}, ... ,\theta _1 \right)\,\, \} 
	\end{align}
	in which $P(N,i)$ is the $i$-permutation of $N$ defined as $P(N,i)=\frac{N!}{\left( N-i \right) !}$.
	The element $A_{ij}$ in PES  is called the \textbf{permutation event}, which is a tuple representing a possible permutation of  $\Theta$,  where $i$ indicates the index for the cardinality of $A_{ij}$ and $j$ denotes the index for the possible permutation.
\end{myDef}


\begin{myDef}[Random permutation set] Given a fixed set of $N$ mutually exclusive and exhaustive elements $\Theta= \{\theta_1, \theta_2, \ldots, \theta_N\}$, its \textbf{random permutation set (RPS)} is a set of pairs defined as follows:
	\begin{align}
	RPS\left( \Theta \right) =\left\{ \left< A,\mathscr{M}\left( A \right) \right> \,\lvert \, A\in PES\left( \Theta \right) \right\} 
	\end{align}
	where $\mathscr{M}$ is called the \textbf{permutation mass function (PMF)},  which is defined as:
	\begin{equation}
	\mathscr{M}: PES(\Theta) \rightarrow [0, 1]
	\end{equation}
	constrained by
	$\mathscr{M}(\emptyset) = 0$ and $ \sum_{A\in PES(\Theta)}{\mathscr{M}(A)}= 1 $.
\end{myDef}


\subsection{Entropy of random permutation set}

For modeling the   uncertainty of the RPS, \cite{deng2021RPSentropy} presented the entropy of random permutation. 

\begin{myDef}[Entropy of random permutation set]
	Let a RPS be denoted as $RPS\left( \Theta \right) =\left\{ \left< A_{ij},\mathscr{M} \left( A_{ij} \right) \right> \,\,\lvert  A_{ij}\in PES\left( \Theta \right) \right\} $, which is defined on  $PES\left( \Theta \right) =\left\{ A_{ij}\,\,\lvert  i=0, ... , N;  j=1, ... ,P\left( N,i \right)  \right\}$. The entropy of this RPS is defined as:
	\begin{align}\label{EntropyRPS}
	H_{RPS}\left( \mathscr{M} \right) =-\sum_{i=1}^N{\sum_{j=1}^{P\left( N,i \right)}{\mathscr{M} \left( A_{ij} \right) \log \left( \frac{\mathscr{M} \left( A_{ij} \right)}{F\left( i \right) -1} \right)}}
	\end{align}
	where  $P(N,i)=\frac{N!}{\left( N-i \right) !}$ is the $i$-permutation of $N$ and $F\left( i \right) =\sum_{k=0}^i{P\left( i,k \right)}=\sum_{k=0}^i{\frac{i!}{\left( i-k \right) !}}$ is the sum from $0$-permutation of $i$ to  $i$-permutation of $i$.
	
\end{myDef}


\section{Maximum entropy of random permutation set}

In this section, the maximum entropy of RPS as well as its PMF condition are respectively presented and proofed.

\subsection{The PMF condition for maximum entropy of RPS}

Let the PES be $PES\left( \Theta \right) =\left\{ A_{ij}\,\,\lvert  i=0, ... , N;  j=1, ... ,P\left( N,i \right)  \right\}$, $P(N,i)$ be as follows $P(N,i)=\frac{N!}{\left( N-i \right) !}$,  and  $F\left( i \right)$ be  defined as  $F\left( i \right) =\sum_{k=0}^i{P\left( i,k \right)}=\sum_{k=0}^i{\frac{i!}{\left( i-k \right) !}}$.  The PMF condition for maximum entropy of RPS is presented as follows.

\begin{theorem}[The PMF condition for maximum entropy of RPS]\label{theorem3}
	The maximum  entropy of RPS happens if and only if the PMF satisfies the following condition
	\begin{equation}
	\mathscr{M} \left( A_{ij} \right) =\frac{F\left( i \right) -1}{\sum_{i=1}^N{\left[ P\left( N,i \right) \left( F\left( i \right) -1 \right) \right]}}
	\end{equation}
\end{theorem}


\begin{myproof}[Proof for Theorem 3]
	
	Let the entropy of RPS be denoted as
	\begin{equation}
	H\left( \mathscr{M} \right) =-\sum_{i=1}^N{\sum_{j=1}^{P\left( N,i \right)}{\mathscr{M} \left( A_{ij} \right) \log_b \left( \frac{\mathscr{M} \left( A_{ij} \right)}{F\left( i \right) -1} \right)}}
	\end{equation}
	which is contrained by
	\begin{equation} \label{m=1}
	\sum_{A_{ij}\in PES(\Theta )}{\mathscr{M} (A_{ij})}=1
	\end{equation}
	where b is the base of logarithmic function. Then  the Lagrange function with Lagrangian multiplier $\lambda$ can be defined as follows:
	\begin{small}
		\begin{equation}
		H_0\left( \mathscr{M} \right) =-\sum_{i=1}^N{\sum_{j=1}^{P\left( N,i \right)}{\mathscr{M} \left( A_{ij} \right) \log _b\left( \frac{\mathscr{M} \left( A_{ij} \right)}{F\left( i \right) -1} \right)}}+\lambda \left( \sum_{A_{ij}\in PES(\Theta )}{\mathscr{M} (A_{ij})}-1 \right) 
		\end{equation}
	\end{small}
	In order to get the maximum of $H\left( \mathscr{M} \right)$, the gradient of $H_{0}\left( \mathscr{M} \right)$ should be equal to $0$. Hence, the gradient can be deduced as follows:
	\begin{align}\label{GRAD}
	\frac{\partial H_0\left( \mathscr{M} \right)}{\partial \mathscr{M} (A_{ij})}=-\log _b\left( \frac{\mathscr{M} (A_{ij})}{F\left( i \right) -1} \right) -\frac{1}{\ln b}+\lambda =0
	\end{align}
	Based on Eq.(\ref{GRAD}), it can be concluded that, with respect to different $i\in\{1,2, ... ,N\}$ and $j\in \{1,2, ... , P(N,i)\}$, 
	$\frac{ \mathscr{M} (A_{ij})}{F\left( i \right) -1}$ is a constant denoted as $C$:
	
	\begin{equation}\label{defC}
	C \triangleq \frac{\mathscr{M} (A_{ij})}{F\left( i \right) -1}
	\end{equation}
	According to Eq.(\ref{defC}) and Eq.(\ref{m=1}), we can get:
	\begin{align}
	\sum_{A_{ij}\in PES(\Theta )}{\mathscr{M} (A_{ij})}&=\sum_{i=1}^N{\sum_{j=1}^{P\left( N,i \right)}{\mathscr{M} \left( A_{ij} \right)}}=\sum_{i=1}^N{\sum_{j=1}^{P\left( N,i \right)}{C*\left[ F\left( i \right) -1 \right]}}
	\notag\\
	&=\sum_{i=1}^N{\left[ P\left( N,i \right) *C*\left( F\left( i \right) -1 \right) \right]}=1
	\end{align}
	so that $C$ can be calculated as 
	\begin{equation}\label{defC2}
	C=\frac{1}{\sum_{i=1}^N{\left[ P\left( N,i \right) \left( F\left( i \right) -1 \right) \right]}}
	\end{equation}
	Based on Eq.(\ref{defC}) and Eq.(\ref{defC2}), we can get this equation:
	\begin{equation}
	\frac{\mathscr{M} (A_{ij})}{F\left( i \right) -1}=\frac{1}{\sum_{i=1}^N{\left[ P\left( N,i \right) \left( F\left( i \right) -1 \right) \right]}}
	\end{equation}
	Therefore, the condition for maximum entropy of RPS can be obtained:
	\begin{equation}\label{distri}
	\mathscr{M} \left( A_{ij} \right) =\frac{F\left( i \right) -1}{\sum_{i=1}^N{\left[ P\left( N,i \right) \left( F\left( i \right) -1 \right) \right]}}
	\end{equation}
	
	~
	\qed
	
\end{myproof}


\subsection{The analytic solution for maximum  entropy of RPS}

Let  $P(N,i)$ be as follows $P(N,i)=\frac{N!}{\left( N-i \right) !}$, and  $F\left( i \right)$ be defined as $F\left( i \right) =\sum_{k=0}^i{P\left( i,k \right)}=\sum_{k=0}^i{\frac{i!}{\left( i-k \right) !}}$.  The analytical solution for the maximum entropy of RPS is detailed as follows.

\begin{theorem}[The analytic solution for maximum  entropy of RPS]\label{theorem4}
	The analytical solution for maximum  entropy of RPS is that
	\begin{equation}
	H_{\max \text{-} RPS}=\log \left( \sum_{i=1}^N{\left[ P\left( N,i \right) \left( F\left( i \right) -1 \right) \right]} \right)  
	\end{equation}
\end{theorem}


\begin{myproof}[Proof for Theorem 4]

	According to \textbf{Theorem \ref{theorem3}}, the maximum entropy of RPS happens when PMF  satisfies   
	Eq.(\ref{distri}). Hence,  the analytic solution  for maximum  entropy of RPS can be calculated by substituting Eq.(\ref{distri})    into Eq.(\ref{EntropyRPS}):
	\begin{small}
		\begin{align}\label{Mexpress}
		H_{\max \text{-} RPS}=-\sum_{i=1}^N{\sum_{j=1}^{P\left( N,i \right)}{\left\{ \frac{F\left( i \right) -1}{\sum_{i=1}^N{\left[ P\left( N,i \right) \left( F\left( i \right) -1 \right) \right]}}\log \left( \frac{1}{\sum_{i=1}^N{\left[ P\left( N,i \right) \left( F\left( i \right) -1 \right) \right]}} \right) \right\}}}
		\end{align}
	\end{small}
	For a certain PES, $N$ is a constant, so that
	$\sum_{i=1}^N{\left[ P\left( N,i \right) \left( F\left( i \right) -1 \right) \right]}$
	is also a constant. Therefore, Eq.(\ref{Mexpress}) can be calculated as:
	\begin{small}
		\begin{align}\label{Mexpress2}
		H_{\max \text{-}RPS}&=\frac{-\log \left( \frac{1}{\sum_{i=1}^N{\left[ P\left( N,i \right) \left( F\left( i \right) -1 \right) \right]}} \right)}{\sum_{i=1}^N{\left[ P\left( N,i \right) \left( F\left( i \right) -1 \right) \right]}}\sum_{i=1}^N{\sum_{j=1}^{P\left( N,i \right)}{\left( F\left( i \right) -1 \right)}}
		\\
		&=\frac{\log \left( \sum_{i=1}^N{\left[ P\left( N,i \right) \left( F\left( i \right) -1 \right) \right]} \right)}{\sum_{i=1}^N{\left[ P\left( N,i \right) \left( F\left( i \right) -1 \right) \right]}}\sum_{i=1}^N{\left[ P\left( N,i \right) \left( F\left( i \right) -1 \right) \right]}
		\\
		&=\log \left( \sum_{i=1}^N{\left[ P\left( N,i \right) \left( F\left( i \right) -1 \right) \right]} \right) 
		\end{align}
	\end{small}
	As a result, the analytic solution for maximum entropy of RPS is obtained.
	
	\qed
\end{myproof}


\section{Numerical examples and discussions}

In this section, some numerical examples are shown to illustrate the presented maximum entropy of RPS and its PMF condition.

\begin{example}
	Assume  there are three balls in a box, whose colors  are red, blue, and green, respectively. With respect to different colors, these balls in the box can be denoted by $\{ R, B, G \}$. Then, we take the following actions:
	
	\begin{enumerate}[\emph{Action} 1:]
		\item Randomly take out \textbf{one ball} from the box. 
		\item Randomly take out \textbf{a number of balls} from the box \textbf{without replacement}. 
		\item Randomly take out \textbf{a number of balls} from the box \textbf{in sequence without replacement}.
	\end{enumerate}
	\begin{figure}[h]
		\centering
		\includegraphics[width=1\linewidth]{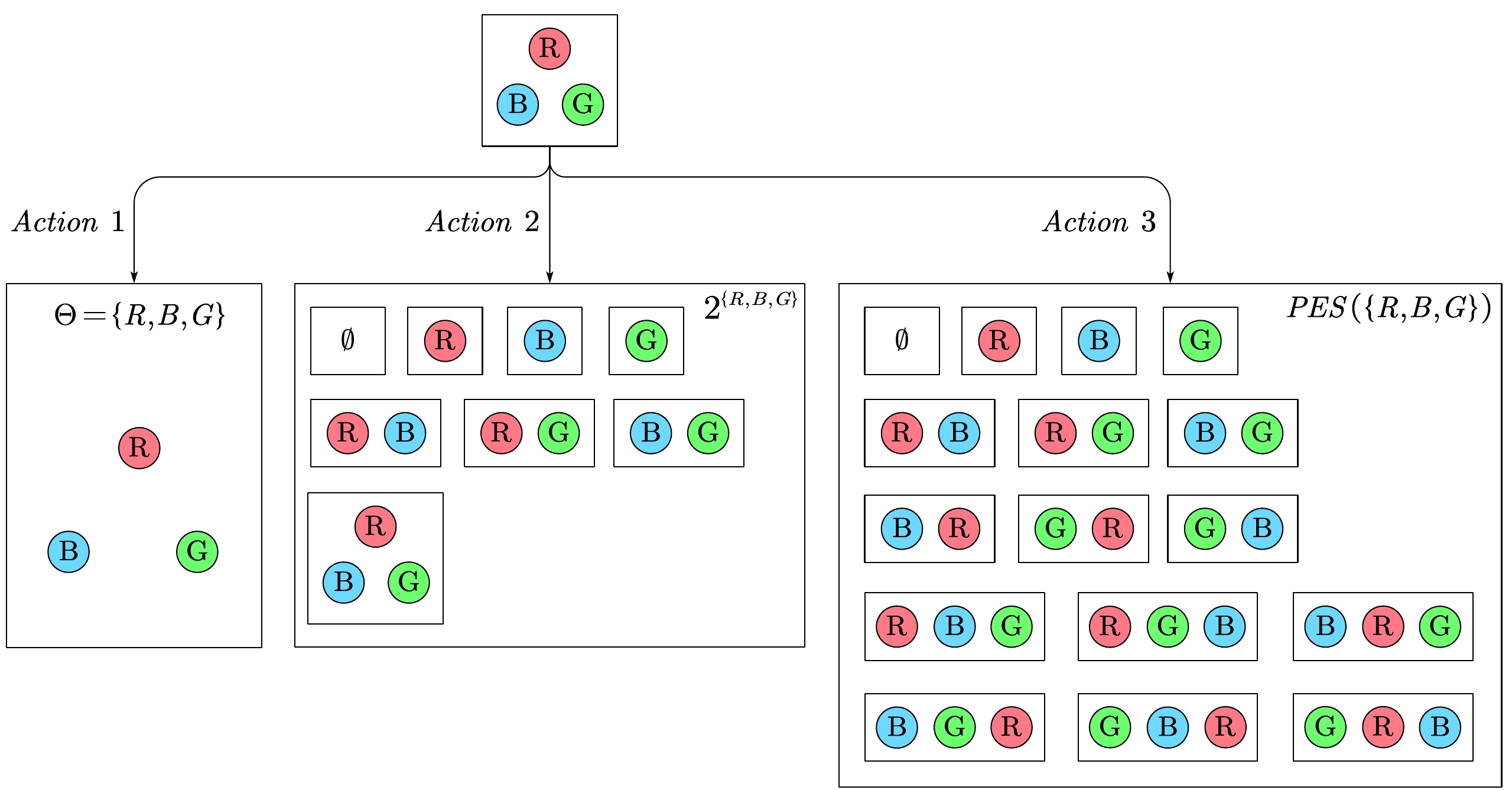}
		\caption{The illustration for all the possible results based on Action 1 to 3}
		\label{fig1}
	\end{figure}

	All the possible results of the three actions can be respectively represented by sample space $\Theta$, power set $2^{\{ R, B, G \}}$, and permutation event space $PES(\{ R, B, G \})$, which are illustrated in Figure~\ref{fig1} and detailed  as follows:

	\begin{align}
	\Theta = \{ R, B, G \}
	\end{align}
	\begin{align}
	2^{\{ R, B, G \}}=\{ \,\,& \emptyset ,\{ R \}  ,\{ B \} ,  \{ G \},
	\\
	&\{ R, B \}  ,\{ R, G \}  ,\{ B, G \}  ,\{ R, B, G \}   \,\, \} \notag
	\end{align}
	\begin{align}
	PES\left( \{ R, B, G \} \right)=\{ \,\, &\emptyset ,\left( R \right)  ,\left( B \right) ,\left( G \right) ,  
	\\
	&\left( R,B \right) ,\left( R,G \right) ,\left( B,G \right) ,
	\notag\\
	&\left( B,R \right) ,\left( G,R \right) ,\left( G,B \right) ,
	\notag\\
	&\left( R,B,G \right) ,\left( R,G,B \right) ,\left( B,R,G \right) ,
	\notag\\
	& \left( B,G,R \right) ,\left( G,B,R \right) ,\left( G,R,B \right) \,\, \} \notag 
	\end{align}
	
	Based on $\Theta$, $2^{\{ R, B, G \}}$, and $PES\left( \{ R, B, G \} \right)$, the probability condition for maximum Shannon entropy~\citep{Shannon1948A}, the mass function condition for maximum Deng entropy~\citep{deng2020uncertainty}, and the PMF condition for maximum entropy of RPS can be respectively obtained:
	\begin{align}
	P:\,\,\,    &P\left( \left\{ R \right\} \right) =P\left( \left\{ B \right\} \right) =P\left( \left\{ G \right\} \right) =0.3333.
	\label{eq1}\\
	m:\,\,\,    &m\left( \left\{ R \right\} \right) =m\left( \left\{ B \right\} \right) =m\left( \left\{ G \right\} \right) =0.0526;
	\label{eq2}\\
	&m\left( \left\{ R,B \right\} \right) =m\left( \left\{ R,G \right\} \right) =m\left( \left\{ B,G \right\} \right) =0.1579;
	\notag\\
	&m\left( \left\{ R,B,G \right\} \right) =0.3684.
	\notag\\
	\mathscr{M}:\,\,\,    &\mathscr{M} \left( R \right) =\mathscr{M} \left( B \right) =\mathscr{M} \left( G \right) =0.0085;
	\label{eq3}\\
	&\mathscr{M} \left( R,B \right) =\mathscr{M} \left( R,G \right) =\mathscr{M} \left( B,G \right) 
	\notag\\
	&=\mathscr{M} \left( B,R \right) =\mathscr{M} \left( G,R \right) =\mathscr{M} \left( G,B \right) =0.0342;
	\notag\\
	&\mathscr{M} \left( R,B,G \right) =\mathscr{M} \left( R,G,B \right) =\mathscr{M} \left( B,R,G \right) 
	\notag\\
	&=\mathscr{M} \left( B,G,R \right) =\mathscr{M} \left( G,B,R \right) =\mathscr{M} \left( G,R,B \right) =0.1282.
	\notag
	\end{align}

\end{example}

	This example illustrate the PMF condition for maximum entropy of RPS. It can be seen in Eq.~(\ref{eq3}) that, with respect to the same cardinality of permutation event,  the PMFs for that permutation events have the same value. For example, the permutation events $(R)$, $(B)$, $(G)$ are of the same value of cardinality. Their corresponding PMFs are the same, which is $0.0085$. In addition, according to Eq.~(\ref{eq3}),  the larger the cardinality of a certain permutation event, the larger the corresponding PMF is.

\begin{example}
	
	Let a set of $N$ mutually exclusive and exhaustive elements be denoted by $\Theta= \{\theta_1, \theta_2, \ldots, \theta_N\}$. Based on $\Theta$, the sample space of maximum Shannon entropy, the power set of maximum Deng entropy, and the PES of maximum entropy of RPS can be respectively indicated by $\Theta$, $2^\Theta$, and $PES(\Theta)$. Hence, their corresponding  maximum  Shannon entropy $H_{\max \text{-} SE}$, maximum  Deng entropy $H_{\max \text{-} DE}$, and maximum entropy of RPS $H_{\max \text{-} RPS}$  can be obtained:
	\begin{small}
		\begin{align}
		H_{\max \text{-} SE} &= \log_2\left(N\right),\\
		H_{\max \text{-} DE} &=   \log_2 \left(\sum_{A \in 2^{\Theta}}(2^{\left\lvert A\right\lvert }-1) \right),\\
		H_{\max \text{-} RPS} &=  \log_2 \left( \sum_{i=1}^N{\left[ P\left( N,i \right) \left( F\left( i \right) -1 \right) \right]} \right).
		\end{align}
	\end{small}
	With different value of $N$, the value of $H_{\max \text{-} SE}$,  $H_{\max \text{-} DE}$, and $H_{\max \text{-} RPS}$ can be calculated, which are shown in Table~\ref{tab1} and Figure~\ref{fig2}. 

	\begin{table}[htbp]
		\centering
		\caption{Maximum  Shannon entropy, maximum  Deng entropy, and maximum entropy of RPS with different $N$ }
		\begin{tabular}{c| ccc}
			\toprule
			$N$ & $H_{\max \text{-} SE}$ & $H_{\max \text{-} DE}$ &  $H_{\max \text{-} RPS}$ \\
			\hline
			1     & 0.0000 & 0.0000 & 0.0000 \\
			2     & 1.0000 & 2.3219 & 3.3219 \\
			3     & 1.5850 & 4.2479 & 6.8704 \\
			4     & 2.0000 & 6.0224 & 10.9278 \\
			5     & 2.3219 & 7.7211 & 15.5406 \\
			6     & 2.5850 & 9.3772 & 20.6691 \\
			7     & 2.8074 & 11.0077 & 26.2495 \\
			8     & 3.0000 & 12.6223 & 32.2231 \\
			9     & 3.1699 & 14.2266 & 38.5424 \\
			10    & 3.3219 & 15.8244 & 45.1699 \\
			\bottomrule
			\multicolumn{4}{l}{Note: the logarithm is based on $2$.}\\
		\end{tabular}%
		\label{tab1}%
	\end{table}%
	
	\begin{figure}[h]
		\centering
		\includegraphics[width=0.7\linewidth]{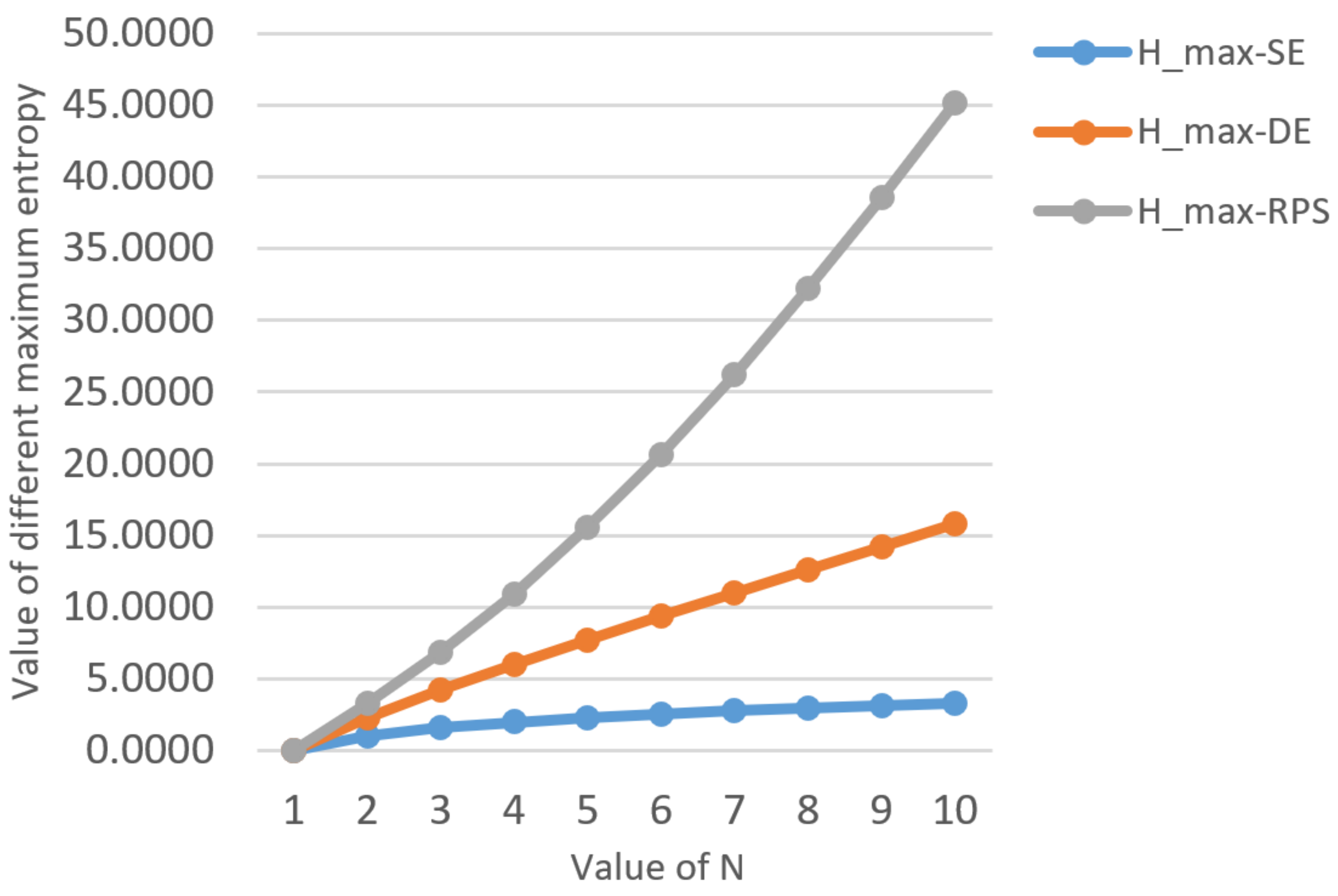}
		\caption{The trend of maximum  Shannon entropy, maximum  Deng entropy, and maximum entropy of RPS with  different $N$}
		\label{fig2}
	\end{figure}

\end{example}

	This example shows that the trend of the maximum entropy of RPS  changing with different value of $N$. In addition, according to Figure~\ref{fig2}, with respect to the same value of $N$,  the  value of maximum entropy of RPS is larger than that of  maximum Shannon entropy or maximum Deng entropy. This means that, with respect to the same number of samples of a certain set,  the uncertainty of a RPS is larger than that of a power set or a sample space, since RPS takes permutation of a certain set into account while sample space and  power set  do not consider that.

\begin{example}
	Given a fixed set of $N$ mutually exclusive and exhaustive elements $\Theta = \{X, Y\}$, its corresponding PES is that
	\begin{align}
	PES(\Theta) = \{ \emptyset,\, (X),\,(Y),\,(X,Y),\,(Y,X)   \}
	\end{align}
	Based on this PES, the maximum entropy of RPS is calculated by:
	\begin{align}
	H_{\max \text{-} RPS} &=  \log_2 \left( \sum_{i=1}^2{\left[ P\left( 2,i \right) \left( F\left( i \right) -1 \right) \right]} \right)
	\\
	&=\log _2\left[ 2*\left( 1+1-1 \right) +2*\left( 1+2+2-1 \right) \right] =3.3219 \notag
	\end{align}
	in which $F(i)$ is defined as $F(i)=\sum_{k=0}^{i}{P(i,k)}$.
	
	Consider the following two scenarios:
	\begin{itemize}
		\item According to \cite{deng2021RPS}, if the order of the element in $PES(\Theta)$ is ignored, $PES(\Theta)$  will degenerate into the power set: $2^\Theta = \{ \emptyset,\, \{X\},\, \{Y\},\, \{X,Y\} \}$. Based on this power set, calculate the maximum Deng entropy
		\begin{align}
		H_{\max \textbf{-}DE} &=\log _2\left( \sum_{A\in 2^{\Theta}}{(}2^{\left\lvert  A \right\lvert }-1) \right) 
		\\
		&=\log _2\left[ 2*\left( 2^1-1 \right) +1*\left( 2^2-1 \right) \right] =2.3219
		\end{align}
		Besides, since the order of the element is ignored, the permutation number $P(N,i)$ degenerates into combinatorial number $C(N,i)$, and $F(i)$ should be calculated as $F(i)=\sum_{k=0}^{i}{C(i,k)}$. Hence, the corresponding maximum entropy of RPS in this scenario can be detailed by:
		\begin{align}
		H_{\max \textbf{-}RPS}&=\log _2\left[ C\left( 2,1 \right) \left( F\left( 1 \right) -1 \right) +C\left( 2,2 \right) \left( F\left( 2 \right) -1 \right) \right] 
		\\
		&=\log _2\left[ 2*\left( 1+1-1 \right) +1*\left( 1+2+1-1 \right) \right] =2.3219 \notag
		\end{align}
		which is the same as $H_{\max \textbf{-}DE}$. As a result, if the order of the element in PES is ignored, the maximum entropy of RPS will degenerate into the maximum  Deng entropy.
		\item According to \cite{deng2021RPS}, if each permutation event is limited to containing just one element,  $PES(\Theta)$  will degenerate into the sample space in probability theory: $\Omega = \{ \{X\},\, \{Y\} \}$. Based on this sample space, calculate the maximum Shannon entropy:
		\begin{align}
		H_{\max \textbf{-}SE} =\log _2\left( 2 \right)=1
		\end{align}
		Additionally, since permutation event is limited to containing just one element, the permutation number $P(N,i)$ degenerates into
		\begin{equation}
		\widetilde{P}\left( N,i \right) \triangleq  \begin{cases}
		1  \,\,   \left( i=0,N\ge 1 \right)\\
		N \,\,     \left( i=1,N\ge 1 \right)\\
		0    \,\, \left( otherwise \right)\\
		\end{cases}  
		\end{equation}
		so that $F(i)$ should be calculated as $F(i)=\sum_{k=0}^{i}{\widetilde{P}(i,k)}=i+1$. Hence, the corresponding maximum entropy of RPS in this scenario is that:
		\begin{align}
		H_{\max \textbf{-}RPS}&=\log _2\left[ \widetilde{P}\left( 2,1 \right)* \left( F\left( 1 \right) -1 \right) +\widetilde{P}\left( 2,2 \right)* \left( F\left( 2 \right) -1 \right) \right] 
		\\
		&=\log _2\left[ 2*\left( 1+1-1 \right) +0*\left( 2+1-1 \right) \right] =1 \notag
		\end{align}
		which is the same as $H_{\max \textbf{-}SE}$. As a result, if each permutation event is limited to containing just one element, the maximum  entropy of RPS degenerates into the maximum  Shannon entropy.
	\end{itemize}

\end{example}
This example shows that the maximum entropy RPS is compatible with the maximum Deng entropy and the maximum Shannon entropy.

\section{Conclusion}

Recently, Random permutation set (RPS)  is proposed, which is a new kind of set considering all the permutation of elements in a certain set.  The entropy of RPS is further presented for modeling the uncertainty of RPS.  However, the maximum  entropy principle of RPS entropy has not been discussed. To address this issue,  in this paper, the maximum entropy of RPS is presented. The main contributions of this paper and some remarks on the maximum entropy of RPS  are summarized as follows:

\begin{itemize}
	\item  The analytical solution for maximum entropy of RPS and its PMF condition are respectively presented and proofed.
	\item Numerical examples are used to illustrate the maximum entropy RPS. The results show  that  the maximum entropy RPS is compatible with the maximum Deng entropy and the maximum Shannon entropy.
	\item When the order of the element in permutation event is ignored, the maximum entropy of RPS will degenerate into the maximum Deng entropy.  When each permutation event is limited to containing just one element, the maximum entropy of RPS will degenerate into the maximum Shannon entropy.
\end{itemize}

\backmatter

%
%
%

\bmhead{Acknowledgments}

The work is partially supported by National Natural Science Foundation of China (Grant No. 61973332); JSPS Invitational Fellowships for Research in Japan (Short-term).

\section*{Declarations}


\begin{itemize}
	\item Funding: The work is partially supported by National Natural Science Foundation of China (Grant No. 61973332); JSPS Invitational Fellowships for Research in Japan (Short-term).
	\item Conflict of interest/Competing interests: All the authors certify that there is no conflict of interest with any individual or organization for this work.
	\item Ethics approval: This article does not 	contain any studies with human participants or animals performed by any of the  authors.
	\item Consent to participate:  Informed consent was obtained from all individual participants included in the study.
	\item Consent for publication: The participant has consented to the submission of the case report to the journal.
	\item Availability of data and materials: All data and materials generated or analysed during this study are included in this article.
	\item Code availability: The code of the current study are available from the corresponding author on reasonable request.
	\item Authors' contributions: All authors contributed to the study conception and design. All authors performed material preparation, data collection and analysis. Jixiang Deng wrote the first draft of the paper.  All authors contributed to the revisions of the paper. All authors read and approved the final manuscript. 
\end{itemize}

\bibliography{mybibfile}


\end{document}